\documentclass[
reprint,
superscriptaddress,
 amsmath,amssymb,
 aps,
]{revtex4-1}

\usepackage{graphicx}
\usepackage{dcolumn}
\usepackage{bm}
\usepackage[mathlines]{lineno}
\usepackage[colorlinks=true,citecolor=blue,linkcolor=red,anchorcolor=green,urlcolor=cyan]{hyperref}
\usepackage{mathrsfs}
\usepackage[usenames, dvipsnames]{xcolor}
\usepackage{float}
\usepackage{ulem}
\usepackage[utf8]{inputenc}
\usepackage{csquotes}

\def\be{\begin{equation}}
\def\ee{\end{equation}}
\def\ba{\begin{eqnarray}}
\def\ea{\end{eqnarray}}

\def\nn{\nonumber}
\def\lf{\left}
\def\rt{\right}

\def\lf{\left}\def\rt{\right}\def\q{\theta}      \def\p {\pi}   \def\d {\delta} \def\f {\phi}  \def\h {\eta}   \def\l {\lambda} \def\z {\zeta} \def\x {\xi}    \def\m {\mu} \def\pd {\partial}\def\p {\pi}   
\def\Q{\Theta}      \def\S {\Sigma}  \def\F {\Phi}  \def\L {\Lambda}    \def\.{\cdot}
\def\math {\mathcal}

\begin{document}

\title{New gedanken experiment on higher-dimensional asymptotically AdS Reissner-Nordström black hole}

\author{Ming Zhang}
\email{mingzhang@jxnu.edu.cn}
\affiliation{Department of Physics, Jiangxi Normal University, Nanchang 330022, China}
\author{Jie Jiang}
\email{jiejiang@mail.bnu.edu.cn (corresponding author)}
\affiliation{Department of Physics, Beijing Normal University, Beijing 100875, China}

\begin{abstract}
Viewing the negative cosmological constant as a dynamical quantity derived from the matter field, we study the weak cosmic censorship conjecture for the higher-dimensional asymptotically AdS Reissner-Nordström black hole. To this end, using stability assumption of matter field perturbation and null energy condition of the matter field, we first derive the first-order and second-order perturbation inequalities containing the variable cosmological constant and its conjugated quantity for the black hole . We prove that the higher-dimensional RN-AdS black hole cannot be destroyed under a second-order approximation of the matter field perturbation process.
\end{abstract}

\maketitle

\section{Introduction}%\label{Introduction}
The weak cosmic censorship conjecture means that the singularity behind the horizon of the black hole formed by gravitational collapse cannot be exposed by any physical process. It also means that the event horizon of the black hole cannot be destroyed by particle matter or general matter. According to the conjecture, to ensure the existence of the event horizon of a black hole with mass $M$, electric charge $Q$ and angular momentum $J$, there must be an inequality
\begin{equation}\label{6015}
\mathcal{F}(M,\, Q,\, J)\geqslant 0.
\end{equation}

It was shown in \cite{WALD1974548} that, the cosmic censorship inequality can not be violated by simply adding particle matter with energy $E$, charge $q$ and angular momentum $l$, even for extreme black hole. That is, we still have
\begin{equation}
\mathcal{F}(M+E,\, Q+q,\, J+l)\geqslant 0
\end{equation}
with the particle matter thrown into the black hole.

For a slightly non-extreme black hole, it was pointed out in \cite{Hubeny:1998ga} that the inequality (\ref{6015}) may be violated by finely tuning the parameters of the particle matter. However, this Hubeny type violation origins from the neglect  of the electromagnetic self-energy and  self-force effects. In fact, one must calculate quadratic-order contribution of the involved parameters of the particle to the energy. Later, by numerical calculations, though not analytical ones, this type of violation for the nearly extreme black hole was proved to be impossible \cite{Zimmerman:2012zu,Barausse:2010ka,Colleoni:2015ena,Barausse:2011vx}.

To test the weak cosmic censorship conjecture for the black hole against any matter perturbation (rather than the particle matter perturbation \cite{Gim:2018axz,Gwak:2017kkt,Zeng:2019jrh,Chen:2019pdj,Shaymatov:2018fmp,Wang:2019jzz,Hu:2019zxr,Han:2019kjr,Yang:2020iat,He:2019kws,Yang:2020czk,Liang:2018wzd,Jiang:2020fgr}), a new version of gedanken experiment, where the self-force effects or finite sizes effects to the energy are regarded,  was raised recently \cite{Sorce:2017dst}. Assuming that the nearly extreme Kerr-Newman black hole can still be stable against linear perturbations of the matter field and null energy condition of the non-electromagnetic contribution to the matter's stress-energy tensor  is not violated, it was explicitly proved that the conjectured inequality (\ref{6015}) can not be violated for the black hole. This new kind of gedanken experiment has been used to test (\ref{6015}) for other black holes, see e.g., \cite{Jiang:2019ige,Jiang:2019vww,Jiang:2019soz,He:2019mqy,Jiang:2020alh,Wang:2020vpn,Liu:2020cji,Shaymatov:2019del,Jiang:2020btc,Jiang:2020mws,Jiang:2019soz,He:2019mqy}. Furthermore, the new gedanken experiment was recently extended from asymptotically flat black hole to asymptotically AdS black hole with a negative cosmological constant \cite{Wang:2019bml}, completing the investigation of the weak cosmic censorship conjecture for the AdS black hole via  particle absorbing process (see e.g., \cite{Zeng:2019jrh,Zeng:2019hux}). The philosophy viewing the cosmological constant  as a dynamical variability  due to the evolution of the system composed by the black hole and arbitrary matter.

In \cite{Ge:2017vun}, it was shown that the weak cosmic censorship conjecture cannot be violated for the nearly extreme higher-dimensional RN black hole with the new gedanken experiment.  In this compact article, wondering whether the Hubeny type violation could occur in the AdS spacetime,  we will use the new gedanken experiment to test the weak cosmic censorship conjecture for asymptotically AdS higher-dimensional RN black hole. We will derive the first-order and second-order perturbation inequalities in Sec. \ref{2616}. Then in Sec. \ref{3061} we will prove that the higher-dimensional RN-AdS black hole can not be destroyed by arbitrary matter and the weak cosmic censorship conjecture is respected. Sec. \ref{4461} is devoted to our conclusion.

\section{First-order and second-order perturbation inequalities}\label{2616}
We provide some preliminaries of the RN-AdS black hole in Einstein-Maxwell gravity in Appendix \ref{ap}.  The first-order variation of the Einstein-Maxwell Lagrangian $\bm{L}$ is
\begin{equation}
	\delta \bm{L}=\bm{E}_{\phi} \delta \phi+d \boldsymbol{\Theta}(\phi, \delta \phi),
\end{equation}
where we use $\delta$ to denote a derivative valued at 0 point of the variation parameter $\lambda$ for the field configurations $\phi$. We have
\begin{equation}\label{eom}
	\boldsymbol{E}(\phi) \delta \phi=-\boldsymbol{\epsilon}\left[\frac{1}{2} T^{a b} \delta g_{a b}+j^{a} \delta A_{a}\right],
\end{equation}
where
\begin{equation}
	T_{a b}=\frac{1}{8 \pi}\left(R_{a b}-\frac{1}{2} R g_{a b}\right)-T_{a b}^{\mathrm{EM}},
\end{equation}
\begin{equation}
 j^{a}=\frac{1}{4 \pi} \nabla_{a} F^{a b}.
\end{equation}
The symplectic potential can be discomposed into gravity part and electromagnetic part,
\begin{equation}
	\boldsymbol{\Theta}_{} = \boldsymbol{\Theta}_{}^\text{GR} + \boldsymbol{\Theta}_{}^\text{EM},
\end{equation}
where
\ba
\begin{aligned}
&\bm{\Theta}_{a_{1}\cdots a_{D-1}}^\text{GR}(\phi, \delta \phi)\\&=\frac{1}{8 \pi} \bm{\epsilon}_{ba_1\cdots a_{D-1}} \left(P_{a}{}^{cbd}\delta\Gamma^{a}_{cd}+\delta g_{bd}\nabla_{a}P^{acbd}\right),\\
&\bm{\Theta}_{a_{1}\cdots a_{D-1}}^\text{EM}(\phi, \delta \phi)=-\frac{1}{4 \pi} \bm{\epsilon}_{ba_{1}\cdots a_{D-1}} F^{bc} \delta A_{c}\,
\end{aligned}
\ea
with
\begin{equation}
P^{abcd}=\frac{1}{2}\left(g^{ac}g^{bd}-g^{ad}g^{bc}\right).
\end{equation}

Associating to a smooth vector field $\zeta^a$ and the field configuration $\phi$ on the spacetime manifold, there is a Noether current $(D-1)$-form
\begin{equation}\label{3281}
\bm{J}_{\zeta}=\boldsymbol{\Theta}\left(\phi, \mathcal{L}_{\zeta} \boldsymbol{\phi}\right)-\zeta \cdot \boldsymbol{L}.
\end{equation}
It has been proved in \cite{Iyer:1995kg} that the Noether current can also be expressed as
\ba\begin{aligned}
\bm{J}_{\zeta}=\boldsymbol{C}_{\zeta}+d \boldsymbol{Q}_{\zeta},
\end{aligned}\ea
where $\bm{Q}_{\zeta}$ is the Noether charge $(D-2)$-form
\begin{equation}
\bm{Q}_\zeta = \bm{Q}_\zeta^\text{GR}+\boldsymbol{Q}_\zeta^\text{EM},
\end{equation}
with
\ba\label{1038}
\begin{aligned}
&\left(\bm{Q}_{\zeta}^{\mathrm{GR}}\right)_{a_{1}\cdots a_{D-2}}\\ &=-\frac{1}{16 \pi} \bm{\epsilon}_{a b a_{1}\cdots a_{D-2}}\left(P^{abcd}\nabla_{c}\z_{d}-2\z_{d}\nabla_{c}P^{abcd}\right),\\
&\left(\bm{Q}_{\zeta}^{\mathrm{EM}}\right)_{a_{1}\cdots a_{D-2}} =-\frac{1}{8 \pi} \bm{\epsilon}_{a b a_{1}\cdots a_{D-2}} F^{ab} A_{c} \zeta^{c}.
\end{aligned}
\ea
$\bm{C}_\zeta = \zeta\cdot \bm{C}$ are constraints of the gravity theory, with
\begin{equation}
\bm{C}_{ba_{2}\cdots a_{D}}=\epsilon_{aa_{2}\cdots a_{D}}\left(T_{b}{}^{a}+A_{b}j^{a}\right).
\end{equation}
We have $\bm{C}=0$ and $d\bm{J}=0$ if $\boldsymbol{E}(\phi)=0$.

The symplectic current can be defined as
\begin{equation}
\boldsymbol{\omega}\left(\phi, \delta_{1} \phi, \delta_{2} \phi\right)=\delta_{1} \boldsymbol{\Theta}\left(\phi, \delta_{2} \phi\right)-\delta_{2} \boldsymbol{\Theta}\left(\phi, \delta_{1} \phi\right)\,,
\end{equation}
the symplectic current can be conserved in condition that $\delta_{1}\bm{E}_{\phi}=\delta_{2}\bm{E}_{\phi}=0$. According to (\ref{3281}), we have the first-order variation of the Noether current $\bm{J}_{\zeta}$ as
\begin{equation}
\delta\bm{J}_{\zeta}=-\zeta\cdot\left(\bm{E}_{\phi}\delta\phi\right)+\omega (\phi,\delta\phi,\mathcal{L}_\zeta \phi )+d\left(\zeta\cdot \bm{\Theta}\right).
\end{equation}

Next, we choose the vector field $\z$ to be a Killing field in the background spacetime. By choosing a proper gauge condition, we can set $\d \z^a=0$. Then we get the first-order perturbation identitiy
\begin{equation}\label{222501}
d\left[\delta \boldsymbol{Q}_{\zeta}-\zeta \cdot \boldsymbol{\Theta}(\phi, \delta \phi)\right]=-\zeta \cdot \boldsymbol{E}_{\phi} \delta \phi-\delta \bm{C}_{\zeta}
\end{equation}
as well as the second-order perturbation identity
\be\begin{aligned}\label{1202}
&d\left[\delta^{2} \boldsymbol{Q}_{\zeta}-\zeta \cdot \delta \boldsymbol{\Theta}(\phi, \delta \phi)\right]\\ &=\boldsymbol{\omega}\left(\phi, \delta \phi, \mathcal{L}_{\zeta} \delta \phi\right)-\zeta \cdot \boldsymbol{E}_{\phi} \delta^{2} \phi-\zeta \cdot \delta \boldsymbol{E}_{\phi} \delta \phi-\delta^{2} \boldsymbol{C}_{\zeta}\,.
\end{aligned}\ee

We now consider that the higher-dimensional RN-AdS black hole is perturbed by the spherically symmetric charged matter field with one-parameter family of field configurations $\phi(\lambda)$. This corresponds to the $\lambda$-dependent perturbation equations of motion
\begin{equation}
\begin{split}
    &R_{a b}(\lambda)-\frac{1}{2} R(\lambda) g_{a b}(\lambda)=8 \pi\left[T_{a b}^{\mathrm{EM}}(\lambda)+T_{a b}(\lambda)\right], \\
    &\nabla_{a}^{(\lambda)} F^{a b}(\lambda)=4 \pi j^{a}(\lambda),
\end{split}
\end{equation}
where
\begin{equation}
T_{ab}^{\text{EM}}(\l)=\frac{1}{4\pi}\left[F_{ac}(\l)F_{b}{}^{c}(\l)-\frac{1}{4}g_{ab}(\l)F_{cd}(\l)F^{cd}(\l)\right],
\end{equation}
is the stress-energy tensor of the electromagnetic field, and $T_{ab}(\l)$ is the stress-energy tensor of non-electromagnetic matter source. In this paper, we assume that the cosmological constant is a dynamical quantity and therefore it should come from the matter source, i.e., we have
\ba\begin{aligned}
T_{ab}=\frac{\Lambda(0)}{8\pi} g_{ab}
\end{aligned}\ea
for the background geometry. Here we denote $\h=\h(\l=0)$ for the background quantity $\h$. The spacetime in this case can be generally described by
\ba\begin{aligned}
ds^2(\l)=-f(r, v, \lambda) d v^{2}+2\m(r,v,\l)drdv+r^2 d\Omega_{D-2},
\end{aligned}\nn\\\ea
where $\m(r,v,0)=1$ and
\ba\begin{aligned}
f(r,v,0)&=f(r)\\
&=1-\frac{2 M}{r^{D-3}}+\frac{Q^{2}}{r^{2D-6}}-\frac{2\L r^{2}}{(D-1)(D-2)}
\end{aligned}\ea
for the background geometry. We suppose that the spacetime can still be described by the higher-dimensional RN-AdS solution at sufficiently late times after the original black hole being perturbed by general spherically symmetric matter source. It indicates that
\ba\label{5681}\begin{aligned}
f(r, v, \lambda)&=f(r, \lambda)\\
=&1-\frac{2 M(\lambda)}{r^{D-3}}+\frac{Q^{2}(\lambda)}{r^{2D-6}}-\frac{2\L(\l) r^{2}}{(D-1)(D-2)},\\
\m(r, b, \l)&=1\,,\quad\quad\quad \bm{A} =-\frac{Q(\lambda)}{r^{D-3}}dv
\end{aligned}\ea
at sufficiently late times.  This supposition is dubbed as the {\it{stability assumption}} \cite{Sorce:2017dst}. Moreover, we also assume that the perturbation vanishes on the bifurcation surface $B$. i.e., $\f(\l)=\f$ on $B$.

We can choose a hyper-surface $\Sigma$ which starts from perturbation-vanishing future event horizon of the nearly extreme higher-dimensional RN-AdS black hole, continues up through the non-vanishing matter source region, and finally becomes space-like as it extends to infinity. Let $\mathcal{H}$ and $\Sigma_1$ individually  be the horizon portion and space-like portion of the Cauchy surface $\Sigma$, we have
\ba\begin{aligned}
\Sigma=\mathcal{H}\cup\Sigma_1.
\end{aligned}\ea
The globally hyperbolic hyper-surface $\Sigma$ terminates at the  $(D-2)-$dimensional bifurcation surface $B$ of the near extreme $D-$dimensional RN-AdS black hole.

After using the Killing vector $\x^{a}=(\partial/\partial v)^{a}$ for the background fields, (\ref{222501}) can be evaluated at $\l=0$ and integrated over $\S$,
\be\begin{aligned}\label{731301}
\int_{\partial\S} \left[\d\bm{Q}_\x-\x\.\bm{Q}(\f,\d\f)\right]+\int_{\S}\x\.\bm{E}_\f\d\f+\int_{\S}\d\bm{C}_\x=0\,.
\end{aligned}\ee

For the first term on the left side of (\ref{731301}), we have
\be\begin{aligned}
&\int_{\partial\S}\left[\d\bm{Q}_\x-\x\.\bm{Q}(\f,\d\f)\right]\\&=\int_{S_c} \left[\d\bm{Q}_\x-\x\.\bm{Q}(\f,\d\f)\right]-\int_{B}\left[\d\bm{Q}_\x-\x\.\bm{Q}(\f,\d\f)\right]\\&=\int_{S_c} \left[\d\bm{Q}^{\text{GR}}_\x-\x\.\bm{Q}^{\text{GR}}(\f,\d\f)\right]\\&\quad+\int_{S_c} \left[\d\bm{Q}^{\text{EM}}_\x-\x\.\bm{Q}^{\text{EM}}(\f,\d\f)\right]\\&=\delta M -V_c \delta P -\frac{1}{8\pi}\int_{S_{c}}\epsilon_{aba_{1}\cdots a_{D-2}}\d F^{ab}A_{c}\x^{c}\\&=\delta \mathcal{M} -V_c \delta P,
\end{aligned}\ee
where we denote $V_c= \Omega_{D-2}\pi r_c^{D-1}/(D-1)$. To avoid the divergence taken by the cosmological constant as a dynamical quantity,  $S_c$ is a sphere with radius $r_c$ replacing  the asymptotic infinity boundary of a isochronous surface $\S_1$ that  can be described by the late time line element of the higher-dimensional RN-AdS black hole. In above we have used the condition that the perturbations vanish on the bifurcation surface $B$. For the second term, we have
\begin{equation}\begin{aligned}
&\int_{\S}\x\.\bm{E}_\f\d\f\\&=\int_{\mathcal{H}}\x\.\bm{E}_\f\d\f+\int_{\S_{1}}\x\.\bm{E}_\f\d\f\\&=\int_{\S_{1}}\x\.\bm{E}_\f\d\f=-\int_{\S_1}\x\.\bm{\epsilon}\left[\frac{1}{2}T^{ab}\d g_{ab}+j^a\d A_a\right]\\&=\frac{\L}{16\p}\int_{\S_1}\x\.\bm{\epsilon}g_{ab}\d g^{ab}\\&=0,
\end{aligned}\end{equation}
where we have used $\x\.\bm{\epsilon}=0$ on $\mathcal{H}$ as well as $j^a(\l)=0$ and $g^{ab}\d g_{ab}=0$ on $\Sigma_1$. For the third term, we have
\begin{equation}
\begin{aligned}
&\int_{\S}\d\bm{C}_\x \\&\quad \\&=\int_{\Sigma_1}\bm{\epsilon}_{aa_{2}\cdots a_{D}}\left[\d T_b{}^a \x^a+A_b\x^b\d j^a\right]\\&\quad+\int_\mathcal{H}\bm{\epsilon}_{aa_{2}\cdots a_{D}}\left[\d T_b{}^a \x^a+A_b\x^b\d j^a\right]\\&=\int_{\Sigma_1}\bm{\epsilon}_{aa_{2}\cdots a_{D}}\d T_b{}^a \x^b+\int_\mathcal{H}\bm{\epsilon}_{aa_{2}\cdots a_{D}}\left[\d T_b{}^a \x^a+A_b\x^b\d j^a\right]\\&=(V_c-V_H)\d P+\int_\math{H}\bm{\epsilon}_{aa_{2}\cdots a_{D}}\d T_b{}^a \x^b\\&\quad-\F_H \d\left[\int_{\math{H}}{\bm{\epsilon}}_{aa_{2}\cdots a_{D}}j^a\right]\\&=(V_c-V_H)\d P+\int_\math{H}{\bm{\epsilon}}_{aa_{2}\cdots a_{D}}\d T_b{}^a \x^b-\F_H \d \mathcal{Q},
\end{aligned}
\end{equation}
where we have used $j^{a}|_{\S}=0$ and denoted $V_H= \Omega_{D-2}\pi r_H^{D-1}/(D-1)$. As a result, we can write the first-order perturbation equality as
\be\begin{aligned}\label{344090}
\d \mathcal{M}-\F_H \d \mathcal{Q}-V_H\d P&=-\int_\math{H}{\epsilon}_{aa_{2}\cdots a_{n}}\d T^{ba} \x_b\\&=\int_\math{H}\bm{\tilde{\epsilon}}\d T^{ab} \x_ak_b\\&\propto  \int_\math{H}\bm{\tilde{\epsilon}}\d T^{ab} k_ak_b,
\end{aligned}\ee
where we have used the condition that the future-directed vector $k^a$ being normal to the horizon and proportional to the killing vector $\x^{a}$. The volume element on the horizon $\bm{\tilde{\epsilon}}$ is defined via the relation $\epsilon_{aa_{2}\cdots a_{D}}=k_{a}\wedge \tilde{\epsilon}_{a_{2}\cdots a_{D}}$. Provided that the {\it{null energy condition}} is respected by the non-electromagnetic stress energy tensor $\d T^{ba}$, i.e., $\int_\math{H}\d T^{ab} k_ak_b\geqslant 0$, we have the first-order perturbation inequality for the higher-dimensional RN-AdS black hole
\be\label{3620}
\d \mathcal{M}-\F_H \d \mathcal{Q}-V_H\d P\geqslant 0.
\ee
When the stree-energy flux of the non-electromagnetic matter vanishes, we can have the optimal perturbation process for the horizon-corruption of the higher-dimensional RN-AdS black hole at first order.

Like the first-order case, the second-order perturbation identity (\ref{1202}) can also be evaluated at $\l=0$ over $\Sigma$ after choosing the Killing vector $\x^{a}$,
\be\begin{aligned}\label{1016x}
&\int_{\partial\S}\d \left[\d \bm{Q}_\x-\x\.\bm{\Q}(\f,\d\f)\right]-\int_{\S}\d\left(\x\.\bm{E}_\f\d\f\right)\\
&+\int_{\S}\d^2\bm{C}_\x-\int_{\mathcal{H}}\boldsymbol{\omega}\left(\phi, \delta \phi, \mathcal{L}_{\xi} \delta \phi\right) \\&-\int_{\S_{1}}\boldsymbol{\omega}\left(\phi, \delta \phi, \mathcal{L}_{\xi} \delta \phi\right) =0\,.
\end{aligned}\ee
Similar to the first-order case, for the first term on the left side, we have
\be\begin{aligned}
&\int_{\partial\S}\d \left[\d \bm{Q}_\x -\x\.\bm{\Q}(\f,\d\f)\right]\\&=\int_{S_c}\d \left[\d \bm{Q}_\x -\x\.\bm{\Q}(\f,\d\f)\right]\\&=\int_{S_c}\frac{(D-2)\sqrt{|g|}}{16\pi}\frac{\delta^{2} f}{r}d\theta_{1}\wedge d\theta_{2}\wedge\cdots d\theta_{D-2}\\&\quad + \int_{S_c} \frac{2(D-3)}{8\p}\frac{\sqrt{|g|}Q\d^2Q}{r^{2D-5}} (d\q)_1\cdots\wedge(d\q)_{D-2}\\&\quad -\int_{S_c}\frac{1}{8\p}\bm{\epsilon}_{aba_{1}\cdots a_{D-2}}\d F^{ab}\d A_e\x^e\\&\quad-\int_{S_c}\frac{1}{8\p}\bm{\epsilon}_{aba_{1}\cdots a_{D-2}}\d^{2} F^{ab} A_e\x^e\\&=\d^2 \mathcal{M}-V_c\d^2P-\int_{S_c}\frac{1}{8\p}\bm{\epsilon}_{aba_{1}\cdots a_{D-2}}\d F^{ab}\d A_e\x^e.
\end{aligned}\ee
It is evident that the second term vanishes. For the third term, we have
\be\begin{aligned}
\int_{\S}\d^2\bm{C}_\x =(V_c-V_H)\d^2 P-\int_\math{H}\bm{\tilde{\epsilon}}\d^2 T_{ab} \x^ak^b-\F_H \d^2 \mathcal{Q}.
\end{aligned}\ee
For the fourth term, resorting to similar result in \cite{Sorce:2017dst}, we get
\be\begin{aligned}
\int_{\mathcal{H}}\boldsymbol{\omega}\left(\phi, \delta \phi, \mathcal{L}_{\xi} \delta \phi\right)=\int_\math{H}\bm{\tilde{\epsilon}}\d^2 T_{ae}^\text{EM} \x^ak^e\,.
\end{aligned}\ee
In order to calculate the last term, we need to use indirect method. We may choose the one-parameter family $\phi^{\text{RA}}(\l)$, whose perturbations satisfy
\begin{equation}\label{4961}
\begin{aligned}
&\mathcal{M}^\text{RA}(\lambda)=\mathcal{M}+\lambda \delta \mathcal{M}^\text{RA}=\mathcal{M}+\lambda \delta \mathcal{M}\,, \\
&\mathcal{Q}^\text{RA}(\lambda)=\mathcal{Q}+\lambda \delta \mathcal{Q}^\text{RA}=\mathcal{Q}+\lambda \delta \mathcal{Q}\,, \\
&\Lambda^\text{RA}(\lambda)=\Lambda+\lambda \delta \Lambda^\text{RA}=\Lambda+\lambda \delta \Lambda\,.
\end{aligned}
\end{equation}
It is natural that $\d^{2}\mathcal{M}=\d^{2}\mathcal{Q}=\d^{2}\L=0$.  Integrating (\ref{1202}) over $\S_{1}$ yields
\be\begin{aligned}\label{1517}
&\int_{\partial\S_{1}}\d \left[\d \bm{Q}_\x^\text{RA}-\x\.\bm{\Q}(\f^\text{RA},\d\f^\text{RA})\right]-\int_{\S_{1}}\d\left(\x\.\bm{E}_\f^\text{RA}\d\f^\text{RA}\right)\\
&+\int_{\S_{1}}\d^2\bm{C}_\x^\text{RA} -\int_{\S_{1}}\boldsymbol{\omega}\left(\phi^\text{RA}, \delta \phi^\text{RA}, \mathcal{L}_{\xi} \delta \phi^\text{RA}\right) =0\,.
\end{aligned}\ee
As the second and third terms vanish in (\ref{1517}), fortunately, we have
\be\begin{aligned}\label{1510}
&\int_{\S_{1}}\boldsymbol{\omega}\left(\phi^\text{RA}, \delta \phi^\text{RA}, \mathcal{L}_{\xi} \delta \phi^\text{RA}\right) \\&=\int_{\partial\S_{1}}\d \left[\d \bm{Q}_\x^\text{RA}-\x\.\bm{\Q}(\f^\text{RA},\d\f^\text{RA})\right]\\&=\int_{S_{c}}\d \left[\d \bm{Q}_\x^\text{RA}-\x\.\bm{\Q}(\f^\text{RA},\d\f^\text{RA})\right]\\&\quad-\int_{B_{1}}\d \left[\d \bm{Q}_\x^\text{RA}-\x\.\bm{\Q}(\f^\text{RA},\d\f^\text{RA})\right]\\&=-\int_{S_c}\frac{1}{8\p}\bm{\epsilon}_{aba_{1}\cdots a_{D-2}}\d F^{ab}\d A_e\x^e\\&\quad +\int_{B_1}\frac{(D-2) \delta Q^2 \left[\prod_{i=1}^{D-3}\sin^{D-2-i}\theta_{i}\right] r^{3-D}}{8 \pi } \prod_{i=1}^{i=D-2}d\theta_{i}.
\end{aligned}\ee
Then we obtain
\be\begin{aligned}\label{1510x}
&\int_{\S_{1}}\boldsymbol{\omega}\left(\phi^\text{RA}, \delta \phi^\text{RA}, \mathcal{L}_{\xi} \delta \phi^\text{RA}\right)+\int_{S_c}\frac{1}{8\p}\bm{\epsilon}_{aba_{1}\cdots a_{D-2}}\d F^{ab}\d A_e\x^e\\&=\frac{\Omega_{D-2}  \left[(D-2) \delta Q^2 r_{H}^{3-D}\right]}{8 \pi }.
\end{aligned}\ee
So we can express the second-order perturbation equality as
\be\begin{aligned}
&\d^2 \mathcal{M}-\F_H\d^2\mathcal{Q}-V_H\d^2P\\&=\frac{\Omega_{D-2}  \left[(D-2) \delta Q^2 r_{H}^{3-D}\right]}{8 \pi }+\int_\math{H}\tilde{\bm{\epsilon}}\lf(\d^2T_{ab}^\text{EM}+\d^2T_{ab}\rt)k^a\x^b.
\end{aligned}\ee

Finally, we can get the second-order inequality for the perturbation
\be\begin{aligned}\label{245090}
\d^2 \mathcal{M}-\F_H\d^2\mathcal{Q}-V_H\d^2P\geqslant \frac{\Omega_{D-2}  \left[(D-2) \delta Q^2 r_{H}^{3-D}\right]}{8 \pi },
\end{aligned}\ee
if the null energy condition for the matter fields
\begin{equation}
\d^2T_{ab}^\text{EM}+\d^2T_{ab}\geqslant 0
\end{equation}
is fulfilled.

\section{New gedanken experiment to destroy a nearly extreme higher-dimensional RN-AdS black hole}\label{3061}
The key of testing the weak cosmic censorship conjecture for the nearly extreme higher-dimensional RN-AdS black hole under the perturbation is to verdict the sign of the perturbed metric function $f(r,\lambda)$, which, for our convenience, can be used to define a perturbation function
\be\label{318490}
h(\l)=f\left(r_m(\l),\l\right),
\ee
where $r_{m}(\l)$ is the extreme value of $f\lf(r,\l\rt)$ and it can be obtained from
\be\begin{aligned}\label{1802}
\pd_rf\lf(r_m(\l),\l\rt)=0\,,
\end{aligned}\ee
so that (\ref{318490}) gives the minimal value of the metric function at late time. We can obtain the mass parameter of the black hole $M$ in terms of the minimal radius $r_{m}$
\be\begin{aligned}
M=\frac{r_m^{-D-1} \left[\left(D^3-6 D^2+11 D-6\right) Q^2 r_m^4+2 \Lambda  r_m^{2 D}\right]}{(D-3) (D-2) (D-1)}.
\end{aligned}\ee
Then we have the differential relation
\begin{equation}\begin{aligned}
\d r_m=&\frac{1}{(D-1) \left[(D-3)^2 (D-2) Q^2 r^4_m-2 \Lambda  r_m^{2 D+1}\right]}\\&\times\left[2   r_m^{2 D}\delta \Lambda -\left(D^3-6 D^2+11 D-6\right)  r_m^{D+2}\delta M\right. \\&\left.+2r_m \left(D^3-6 D^2+11 D-6\right) Q r_m^4  \delta Q\right].
\end{aligned}\end{equation}
Expanding $h(\lambda)$ in terms of $\lambda$ to second-order level, we have
\begin{equation}\label{546090}
h(\l)=h_{0}+\l h_{1}+\l^{2}(h_{21}+h_{22})+\mathcal{O}(\l^{2}),
\end{equation}
where
\begin{equation}
\begin{aligned}
h_{0} =1-\frac{2 \Lambda  r_m^2}{D^2-5 D+6}-Q^2 r_m^{6-2 D},
\end{aligned}
\end{equation}
\begin{equation}
\begin{aligned}
h_{1} =-\frac{2  r_m^2  \delta \Lambda}{D^2-3 D+2}+2  Q r_m^{6-2 D}\delta Q-2  r_m^{3-D}\delta M,
\end{aligned}
\end{equation}
\begin{equation}
\begin{aligned}
h_{21} =&Q r_m^{4-2 D} \left[(D-3)^2  Q\delta r_m^2+ r_m^2\delta ^2 Q\right]- r_m^{3-D}\delta ^2 M\\&-\frac{2 (D-1)  \Lambda \delta r_m^2 +  r_m^2 \delta ^2 \Lambda}{(D-2) (D-1)},
\end{aligned}
\end{equation}
\begin{equation}
\begin{aligned}
h_{22} =& r_m^{5-2 D} \left[ r_m \delta Q-4 (D-3)  Q\delta r_m\right]\delta Q\\&+2 (D-3)  r_m^{2-D}\delta r_m  \delta M-\frac{4r_m \delta \Lambda  \delta r_m} {D^2-3 D+2}.
\end{aligned}
\end{equation}

With a similar consideration of \cite{Sorce:2017dst}, we consider that the black hole approaches extreme geometry, with
\begin{equation}
r_{m}=(1-\epsilon)r_{H},
\end{equation}
where $\epsilon\to 0$ is in agreement with the extra matter fields' first-order perturbation,  i.e., it is of the same order as $\l$. Note that $\epsilon$ is defined in the background geometry and therefore it is independent of $\l$. Not loss of generality, it is not difficult to deduce
\be
f'\lf(r_H\rt)=\epsilon r_H f''\lf(r_H\rt)
\ee
and
\be\begin{aligned}
f(r_m)&=f\lf(r_H(1-\epsilon)\rt)\\
&\simeq -\epsilon r_H f'(r_H)+\frac{\epsilon^2 r_H^2}{2} f''(r_H)\\
&=-\frac{1}{2}\epsilon^2 r_H^2 f''(r_H)\,.
\end{aligned}\ee

Moreover, considering the zero-order approximation of $\epsilon$, we have
\be\begin{aligned}
\L&=\frac{1}{2} (D-3) (D-2) r_m^{-2 (D+1)} \left(r_m^{2 D}-Q^2 r_m^6\right)\,,\\ M&=\frac{r_m^{-D-3} \left((D-2) Q^2 r_m^6+r_m^{2 D}\right)}{D-1}\,,
\end{aligned}\ee
which can be derived from $f((1+\epsilon)r_m)=0$ and $f'(r_m)=0$. After using the first-order perturbation inequality (\ref{3620}) and the second-order perturbation inequality (\ref{245090}), (\ref{546090}) can be further reduced, as
\be\begin{aligned}\label{4302}
h(\l)&\leqslant \frac{\mathcal{X}(\l)^2 }{4 (D-3) r_m^{-2 (D+1)} \left((2-D) Q^2 r_m^6+r_m^{2 D}\right)}+\mathcal{O}(\l^{2})\\&=-\frac{\mathcal{X}(\l)^2}{2f^{\prime\prime}(r_{m})}+\mathcal{O}(\l^{2}),
\end{aligned}\ee
where $\mathcal{X}(\l)$ is a tedious normal expression, we here will not explicitly show it. As $f(r_{m})\leqslant f(r)$, or in other words, $f(r_{m})$ is the minimum value of the metric function $f(r)$ for the higher-dimensional RN-AdS black hole, we have  $f^{\prime\prime}(r_{m})>0$. Then (\ref{4302}) implies
\be
h(\l)\leqslant 0.
\ee
This result tells us that, to the level of second-order approximation of the perturbation from the extra spherically symmetric matter field, which affects the mass, electric charge as well as the cosmological constant of the higher-dimensional RN-AdS black hole, the event horizon of the nearly extreme black hole cannot be deprived and the weak cosmic censorship conjecture can be roundly respected.

\section{Conclusion}\label{4461}
In this article, supposing that the stress-energy tensors of the non-electromagnetic matters do not violate the null energy condition, and the nearly extreme higher-dimensional RN-AdS black hole comes to be linearly stable at late times under the perturbation of the matter, we derived the first-order perturbation and the second-order perturbation inequalities. We then proved that the weak cosmic censorship conjecture for the higher-dimensional RN-AdS black hole cannot be violated.

There are two kinds of investigations we can do further. The first one is to study that whether the weak cosmic censorship conjecture can be violated by considering higher-order approximations, though it is highly unlikely \cite{Figueras:2017zwa}. The second one is to consider the new gedanken experiment on asymptotically AdS rotating black hole.

\section*{Acknowledgements}
This work is supported by the Initial Research Foundation of Jiangxi Normal University with Grant No. 12020023  and  the National Natural Science Foundation of China with Grant No. 11675015.

\appendix
\section{A brief review of higher-dimensional Reissner-Nordström black hole in anti-de sitter background}\label{ap}
The action of the Einstein-Maxwell gravity with a cosmological constant are
\begin{equation}\label{1810}
\boldsymbol{L}=\frac{ \boldsymbol{\epsilon}}{16 \pi}\left(R-2 \Lambda-F_{a b} F^{a b}\right),
\end{equation}
where $\bm{\epsilon}$ is the volume element, $R$ is the Ricci scalar, $\Lambda$ is the negative cosmological constant, $\bm{F}$ is electromagnetic field strength. The equations of motion is
\begin{equation}
G_{ab}=8\pi \left(T_{ab}^{\text{EM}}+T_{ab}\right),
\end{equation}
\begin{equation}
\nabla_{a}F^{ab}=4\pi j^{b},
\end{equation}
where
\begin{equation}
T_{ab}^{\text{EM}}=\frac{1}{4\pi}\left(F_{ac}F_{b}{}^{c}-\frac{1}{4}g_{ab}F_{cd}F^{cd}\right)
\end{equation}
is the stress energy tensor for the electromagnetic field,
\begin{equation}
T_{a b}=\frac{\Lambda}{8\pi} g_{ab}
\end{equation}
is the stress energy tensor for the matter field, $G_{ab}$ is the Einstein tensor and $j^{a}$ is the electric current for the matter field.

 The $D$-dimensional RN-AdS black hole solution corresponding to the equations of motion is
\begin{equation} \label{4812}
\begin{aligned}
	d s^{2} &= - f(r)dt^2 +f^{-1}(r)dr^{2} + r^2 d\Omega_{D-2}, \\
    \boldsymbol{A} &=-\frac{Q}{r^{D-3}}dt,
\end{aligned}
\end{equation}
where
\begin{equation}
f(r) =1-\frac{2M}{r^{D-3}}+\frac{Q^2}{r^{2D-6}}-\frac{2\L r^2}{(D-1)(D-2)},
\end{equation}
\begin{equation}
d\Omega_{D-2}=\sum^{D-2}_{i=1}\left( \prod_{j=1}^{i} \sin^2\theta_{j-1}\right)d\theta_i^2,\quad \theta_0\equiv\frac{\pi}{2}.
\end{equation}
We can introduce the Eddington-Finkelstein coordinate
\be
v=t+\int\frac{dt}{f(r)},
\ee
then the solution can be written as
\begin{equation} \label{4812x}
\begin{aligned}
d s^{2} &= - f(r)d\nu^2 + 2drd\nu + r^2 d\Omega_{D-2},\\
\boldsymbol{A} &=-\frac{Q}{r^{D-3}}d\nu.
\end{aligned}
\end{equation}

The mass, electric charge, electric potential, cosmological constant, thermodynamic pressure are \cite{Kubiznak:2012wp}
\begin{equation}
\mathcal{M}=\frac{(D-2)\Omega_{D-2}}{8\pi}M,
\end{equation}
\begin{equation}
\mathcal{Q}=\frac{(D-2)\Omega_{D-2}}{8\pi}Q,
\end{equation}
\begin{equation}
\Phi_H=\frac{Q}{r_H^{D-3}},
\end{equation}
\begin{equation}
P=-\frac{\Lambda}{8\pi},
\end{equation}
\begin{equation}
V=\frac{\Omega_{D-2}}{D-1}r_H^{D-1},
\end{equation}
where $\Omega_{D-2}=2\pi^{(D-1)/2}/\Gamma[(D-1)/2]$ is the volume of the $(D-2)$-dimensional sphere.


\begin{thebibliography}{36}%
\makeatletter
\providecommand \@ifxundefined [1]{%
 \@ifx{#1\undefined}
}%
\providecommand \@ifnum [1]{%
 \ifnum #1\expandafter \@firstoftwo
 \else \expandafter \@secondoftwo
 \fi
}%
\providecommand \@ifx [1]{%
 \ifx #1\expandafter \@firstoftwo
 \else \expandafter \@secondoftwo
 \fi
}%
\providecommand \natexlab [1]{#1}%
\providecommand \enquote  [1]{``#1''}%
\providecommand \bibnamefont  [1]{#1}%
\providecommand \bibfnamefont [1]{#1}%
\providecommand \citenamefont [1]{#1}%
\providecommand \href@noop [0]{\@secondoftwo}%
\providecommand \href [0]{\begingroup \@sanitize@url \@href}%
\providecommand \@href[1]{\@@startlink{#1}\@@href}%
\providecommand \@@href[1]{\endgroup#1\@@endlink}%
\providecommand \@sanitize@url [0]{\catcode `\\12\catcode `\$12\catcode
  `\&12\catcode `\#12\catcode `\^12\catcode `\_12\catcode `\%12\relax}%
\providecommand \@@startlink[1]{}%
\providecommand \@@endlink[0]{}%
\providecommand \url  [0]{\begingroup\@sanitize@url \@url }%
\providecommand \@url [1]{\endgroup\@href {#1}{\urlprefix }}%
\providecommand \urlprefix  [0]{URL }%
\providecommand \Eprint [0]{\href }%
\providecommand \doibase [0]{http://dx.doi.org/}%
\providecommand \selectlanguage [0]{\@gobble}%
\providecommand \bibinfo  [0]{\@secondoftwo}%
\providecommand \bibfield  [0]{\@secondoftwo}%
\providecommand \translation [1]{[#1]}%
\providecommand \BibitemOpen [0]{}%
\providecommand \bibitemStop [0]{}%
\providecommand \bibitemNoStop [0]{.\EOS\space}%
\providecommand \EOS [0]{\spacefactor3000\relax}%
\providecommand \BibitemShut  [1]{\csname bibitem#1\endcsname}%
\let\auto@bib@innerbib\@empty
%</preamble>
\bibitem [{\citenamefont {Wald}(1974)}]{WALD1974548}%
  \BibitemOpen
  \bibfield  {author} {\bibinfo {author} {\bibfnamefont {R.}~\bibnamefont
  {Wald}},\ }\href {\doibase 10.1016/0003-4916(74)90125-0} {\bibfield
  {journal} {\bibinfo  {journal} {Annals of Physics}\ }\textbf {\bibinfo
  {volume} {82}},\ \bibinfo {pages} {548 } (\bibinfo {year}
  {1974})}\BibitemShut {NoStop}%
\bibitem [{\citenamefont {Hubeny}(1999)}]{Hubeny:1998ga}%
  \BibitemOpen
  \bibfield  {author} {\bibinfo {author} {\bibfnamefont {V.~E.}\ \bibnamefont
  {Hubeny}},\ }\href {\doibase 10.1103/PhysRevD.59.064013} {\bibfield
  {journal} {\bibinfo  {journal} {Phys. Rev.}\ }\textbf {\bibinfo {volume}
  {D59}},\ \bibinfo {pages} {064013} (\bibinfo {year} {1999})},\ \Eprint
  {http://arxiv.org/abs/gr-qc/9808043} {arXiv:gr-qc/9808043 [gr-qc]}
  \BibitemShut {NoStop}%
%%CITATION = GR-QC/9808043;%%
\bibitem [{\citenamefont {Zimmerman}\ \emph {et~al.}(2013)\citenamefont
  {Zimmerman}, \citenamefont {Vega}, \citenamefont {Poisson},\ and\
  \citenamefont {Haas}}]{Zimmerman:2012zu}%
  \BibitemOpen
  \bibfield  {author} {\bibinfo {author} {\bibfnamefont {P.}~\bibnamefont
  {Zimmerman}}, \bibinfo {author} {\bibfnamefont {I.}~\bibnamefont {Vega}},
  \bibinfo {author} {\bibfnamefont {E.}~\bibnamefont {Poisson}}, \ and\
  \bibinfo {author} {\bibfnamefont {R.}~\bibnamefont {Haas}},\ }\href {\doibase
  10.1103/PhysRevD.87.041501} {\bibfield  {journal} {\bibinfo  {journal} {Phys.
  Rev.}\ }\textbf {\bibinfo {volume} {D87}},\ \bibinfo {pages} {041501}
  (\bibinfo {year} {2013})},\ \Eprint {http://arxiv.org/abs/1211.3889}
  {arXiv:1211.3889 [gr-qc]} \BibitemShut {NoStop}%
%%CITATION = ARXIV:1211.3889;%%
\bibitem [{\citenamefont {Barausse}\ \emph {et~al.}(2010)\citenamefont
  {Barausse}, \citenamefont {Cardoso},\ and\ \citenamefont
  {Khanna}}]{Barausse:2010ka}%
  \BibitemOpen
  \bibfield  {author} {\bibinfo {author} {\bibfnamefont {E.}~\bibnamefont
  {Barausse}}, \bibinfo {author} {\bibfnamefont {V.}~\bibnamefont {Cardoso}}, \
  and\ \bibinfo {author} {\bibfnamefont {G.}~\bibnamefont {Khanna}},\ }\href
  {\doibase 10.1103/PhysRevLett.105.261102} {\bibfield  {journal} {\bibinfo
  {journal} {Phys. Rev. Lett.}\ }\textbf {\bibinfo {volume} {105}},\ \bibinfo
  {pages} {261102} (\bibinfo {year} {2010})},\ \Eprint
  {http://arxiv.org/abs/1008.5159} {arXiv:1008.5159 [gr-qc]} \BibitemShut
  {NoStop}%
%%CITATION = ARXIV:1008.5159;%%
\bibitem [{\citenamefont {Colleoni}\ \emph {et~al.}(2015)\citenamefont
  {Colleoni}, \citenamefont {Barack}, \citenamefont {Shah},\ and\ \citenamefont
  {van~de Meent}}]{Colleoni:2015ena}%
  \BibitemOpen
  \bibfield  {author} {\bibinfo {author} {\bibfnamefont {M.}~\bibnamefont
  {Colleoni}}, \bibinfo {author} {\bibfnamefont {L.}~\bibnamefont {Barack}},
  \bibinfo {author} {\bibfnamefont {A.~G.}\ \bibnamefont {Shah}}, \ and\
  \bibinfo {author} {\bibfnamefont {M.}~\bibnamefont {van~de Meent}},\ }\href
  {\doibase 10.1103/PhysRevD.92.084044} {\bibfield  {journal} {\bibinfo
  {journal} {Phys. Rev. D}\ }\textbf {\bibinfo {volume} {92}},\ \bibinfo
  {pages} {084044} (\bibinfo {year} {2015})},\ \Eprint
  {http://arxiv.org/abs/1508.04031} {arXiv:1508.04031 [gr-qc]} \BibitemShut
  {NoStop}%
\bibitem [{\citenamefont {Barausse}\ \emph {et~al.}(2011)\citenamefont
  {Barausse}, \citenamefont {Cardoso},\ and\ \citenamefont
  {Khanna}}]{Barausse:2011vx}%
  \BibitemOpen
  \bibfield  {author} {\bibinfo {author} {\bibfnamefont {E.}~\bibnamefont
  {Barausse}}, \bibinfo {author} {\bibfnamefont {V.}~\bibnamefont {Cardoso}}, \
  and\ \bibinfo {author} {\bibfnamefont {G.}~\bibnamefont {Khanna}},\ }\href
  {\doibase 10.1103/PhysRevD.84.104006} {\bibfield  {journal} {\bibinfo
  {journal} {Phys. Rev. D}\ }\textbf {\bibinfo {volume} {84}},\ \bibinfo
  {pages} {104006} (\bibinfo {year} {2011})},\ \Eprint
  {http://arxiv.org/abs/1106.1692} {arXiv:1106.1692 [gr-qc]} \BibitemShut
  {NoStop}%
\bibitem [{\citenamefont {Gim}\ and\ \citenamefont {Gwak}(2019)}]{Gim:2018axz}%
  \BibitemOpen
  \bibfield  {author} {\bibinfo {author} {\bibfnamefont {Y.}~\bibnamefont
  {Gim}}\ and\ \bibinfo {author} {\bibfnamefont {B.}~\bibnamefont {Gwak}},\
  }\href {\doibase 10.1016/j.physletb.2019.05.039} {\bibfield  {journal}
  {\bibinfo  {journal} {Phys. Lett.}\ }\textbf {\bibinfo {volume} {B794}},\
  \bibinfo {pages} {122} (\bibinfo {year} {2019})},\ \Eprint
  {http://arxiv.org/abs/1808.05943} {arXiv:1808.05943 [gr-qc]} \BibitemShut
  {NoStop}%
%%CITATION = ARXIV:1808.05943;%%
\bibitem [{\citenamefont {Gwak}(2017)}]{Gwak:2017kkt}%
  \BibitemOpen
  \bibfield  {author} {\bibinfo {author} {\bibfnamefont {B.}~\bibnamefont
  {Gwak}},\ }\href {\doibase 10.1007/JHEP11(2017)129} {\bibfield  {journal}
  {\bibinfo  {journal} {JHEP}\ }\textbf {\bibinfo {volume} {11}},\ \bibinfo
  {pages} {129} (\bibinfo {year} {2017})},\ \Eprint
  {http://arxiv.org/abs/1709.08665} {arXiv:1709.08665 [gr-qc]} \BibitemShut
  {NoStop}%
%%CITATION = ARXIV:1709.08665;%%
\bibitem [{\citenamefont {Zeng}\ \emph
  {et~al.}(2019{\natexlab{a}})\citenamefont {Zeng}, \citenamefont {Han},\ and\
  \citenamefont {Chen}}]{Zeng:2019jrh}%
  \BibitemOpen
  \bibfield  {author} {\bibinfo {author} {\bibfnamefont {X.-X.}\ \bibnamefont
  {Zeng}}, \bibinfo {author} {\bibfnamefont {Y.-W.}\ \bibnamefont {Han}}, \
  and\ \bibinfo {author} {\bibfnamefont {D.-Y.}\ \bibnamefont {Chen}},\ }\href
  {\doibase 10.1088/1674-1137/43/10/105104} {\bibfield  {journal} {\bibinfo
  {journal} {Chin. Phys.}\ }\textbf {\bibinfo {volume} {C43}},\ \bibinfo
  {pages} {105104} (\bibinfo {year} {2019}{\natexlab{a}})},\ \Eprint
  {http://arxiv.org/abs/1901.08915} {arXiv:1901.08915 [gr-qc]} \BibitemShut
  {NoStop}%
%%CITATION = ARXIV:1901.08915;%%
\bibitem [{\citenamefont {Chen}(2019)}]{Chen:2019pdj}%
  \BibitemOpen
  \bibfield  {author} {\bibinfo {author} {\bibfnamefont {D.}~\bibnamefont
  {Chen}},\ }\href {\doibase 10.1140/epjc/s10052-019-6874-5} {\bibfield
  {journal} {\bibinfo  {journal} {Eur. Phys. J.}\ }\textbf {\bibinfo {volume}
  {C79}},\ \bibinfo {pages} {353} (\bibinfo {year} {2019})},\ \Eprint
  {http://arxiv.org/abs/1902.06489} {arXiv:1902.06489 [hep-th]} \BibitemShut
  {NoStop}%
%%CITATION = ARXIV:1902.06489;%%
\bibitem [{\citenamefont {Shaymatov}\ \emph {et~al.}(2019)\citenamefont
  {Shaymatov}, \citenamefont {Dadhich},\ and\ \citenamefont
  {Ahmedov}}]{Shaymatov:2018fmp}%
  \BibitemOpen
  \bibfield  {author} {\bibinfo {author} {\bibfnamefont {S.}~\bibnamefont
  {Shaymatov}}, \bibinfo {author} {\bibfnamefont {N.}~\bibnamefont {Dadhich}},
  \ and\ \bibinfo {author} {\bibfnamefont {B.}~\bibnamefont {Ahmedov}},\ }\href
  {\doibase 10.1140/epjc/s10052-019-7088-6} {\bibfield  {journal} {\bibinfo
  {journal} {Eur. Phys. J. C}\ }\textbf {\bibinfo {volume} {79}},\ \bibinfo
  {pages} {585} (\bibinfo {year} {2019})},\ \Eprint
  {http://arxiv.org/abs/1809.10457} {arXiv:1809.10457 [gr-qc]} \BibitemShut
  {NoStop}%
\bibitem [{\citenamefont {Wang}\ \emph {et~al.}(2019)\citenamefont {Wang},
  \citenamefont {Wu},\ and\ \citenamefont {Yang}}]{Wang:2019jzz}%
  \BibitemOpen
  \bibfield  {author} {\bibinfo {author} {\bibfnamefont {P.}~\bibnamefont
  {Wang}}, \bibinfo {author} {\bibfnamefont {H.}~\bibnamefont {Wu}}, \ and\
  \bibinfo {author} {\bibfnamefont {H.}~\bibnamefont {Yang}},\ }\href {\doibase
  10.1140/epjc/s10052-019-7090-z} {\bibfield  {journal} {\bibinfo  {journal}
  {Eur. Phys. J.}\ }\textbf {\bibinfo {volume} {C79}},\ \bibinfo {pages} {572}
  (\bibinfo {year} {2019})}\BibitemShut {NoStop}%
%%CITATION = EPHJA,C79,572;%%
\bibitem [{\citenamefont {Hu}\ \emph {et~al.}(2019)\citenamefont {Hu},
  \citenamefont {Liu}, \citenamefont {Kuang},\ and\ \citenamefont
  {Yue}}]{Hu:2019zxr}%
  \BibitemOpen
  \bibfield  {author} {\bibinfo {author} {\bibfnamefont {S.-Q.}\ \bibnamefont
  {Hu}}, \bibinfo {author} {\bibfnamefont {B.}~\bibnamefont {Liu}}, \bibinfo
  {author} {\bibfnamefont {X.-M.}\ \bibnamefont {Kuang}}, \ and\ \bibinfo
  {author} {\bibfnamefont {R.-H.}\ \bibnamefont {Yue}},\ }\href@noop {} {\
  (\bibinfo {year} {2019})},\ \Eprint {http://arxiv.org/abs/1910.04437}
  {arXiv:1910.04437 [gr-qc]} \BibitemShut {NoStop}%
%%CITATION = ARXIV:1910.04437;%%
\bibitem [{\citenamefont {Han}\ \emph {et~al.}(2019)\citenamefont {Han},
  \citenamefont {Zeng},\ and\ \citenamefont {Hong}}]{Han:2019kjr}%
  \BibitemOpen
  \bibfield  {author} {\bibinfo {author} {\bibfnamefont {Y.-W.}\ \bibnamefont
  {Han}}, \bibinfo {author} {\bibfnamefont {X.-X.}\ \bibnamefont {Zeng}}, \
  and\ \bibinfo {author} {\bibfnamefont {Y.}~\bibnamefont {Hong}},\ }\href
  {\doibase 10.1140/epjc/s10052-019-6771-y} {\bibfield  {journal} {\bibinfo
  {journal} {Eur. Phys. J. C}\ }\textbf {\bibinfo {volume} {79}},\ \bibinfo
  {pages} {252} (\bibinfo {year} {2019})},\ \Eprint
  {http://arxiv.org/abs/1901.10660} {arXiv:1901.10660 [hep-th]} \BibitemShut
  {NoStop}%
\bibitem [{\citenamefont {Yang}\ \emph
  {et~al.}(2020{\natexlab{a}})\citenamefont {Yang}, \citenamefont {Chen},
  \citenamefont {Wan}, \citenamefont {Wei},\ and\ \citenamefont
  {Liu}}]{Yang:2020iat}%
  \BibitemOpen
  \bibfield  {author} {\bibinfo {author} {\bibfnamefont {S.-J.}\ \bibnamefont
  {Yang}}, \bibinfo {author} {\bibfnamefont {J.}~\bibnamefont {Chen}}, \bibinfo
  {author} {\bibfnamefont {J.-J.}\ \bibnamefont {Wan}}, \bibinfo {author}
  {\bibfnamefont {S.-W.}\ \bibnamefont {Wei}}, \ and\ \bibinfo {author}
  {\bibfnamefont {Y.-X.}\ \bibnamefont {Liu}},\ }\href {\doibase
  10.1103/PhysRevD.101.064048} {\bibfield  {journal} {\bibinfo  {journal}
  {Phys. Rev. D}\ }\textbf {\bibinfo {volume} {101}},\ \bibinfo {pages}
  {064048} (\bibinfo {year} {2020}{\natexlab{a}})},\ \Eprint
  {http://arxiv.org/abs/2001.03106} {arXiv:2001.03106 [gr-qc]} \BibitemShut
  {NoStop}%
\bibitem [{\citenamefont {He}\ \emph {et~al.}(2020)\citenamefont {He},
  \citenamefont {Li},\ and\ \citenamefont {Hu}}]{He:2019kws}%
  \BibitemOpen
  \bibfield  {author} {\bibinfo {author} {\bibfnamefont {K.-J.}\ \bibnamefont
  {He}}, \bibinfo {author} {\bibfnamefont {G.-P.}\ \bibnamefont {Li}}, \ and\
  \bibinfo {author} {\bibfnamefont {X.-Y.}\ \bibnamefont {Hu}},\ }\href
  {\doibase 10.1140/epjc/s10052-020-7669-4} {\bibfield  {journal} {\bibinfo
  {journal} {Eur. Phys. J. C}\ }\textbf {\bibinfo {volume} {80}},\ \bibinfo
  {pages} {209} (\bibinfo {year} {2020})},\ \Eprint
  {http://arxiv.org/abs/1909.09956} {arXiv:1909.09956 [hep-th]} \BibitemShut
  {NoStop}%
\bibitem [{\citenamefont {Yang}\ \emph
  {et~al.}(2020{\natexlab{b}})\citenamefont {Yang}, \citenamefont {Wan},
  \citenamefont {Chen}, \citenamefont {Yang},\ and\ \citenamefont
  {Wang}}]{Yang:2020czk}%
  \BibitemOpen
  \bibfield  {author} {\bibinfo {author} {\bibfnamefont {S.-J.}\ \bibnamefont
  {Yang}}, \bibinfo {author} {\bibfnamefont {J.-J.}\ \bibnamefont {Wan}},
  \bibinfo {author} {\bibfnamefont {J.}~\bibnamefont {Chen}}, \bibinfo {author}
  {\bibfnamefont {J.}~\bibnamefont {Yang}}, \ and\ \bibinfo {author}
  {\bibfnamefont {Y.-Q.}\ \bibnamefont {Wang}},\ }\href@noop {} {\  (\bibinfo
  {year} {2020}{\natexlab{b}})},\ \Eprint {http://arxiv.org/abs/2004.07934}
  {arXiv:2004.07934 [gr-qc]} \BibitemShut {NoStop}%
\bibitem [{\citenamefont {Liang}\ \emph {et~al.}(2019)\citenamefont {Liang},
  \citenamefont {Wei},\ and\ \citenamefont {Liu}}]{Liang:2018wzd}%
  \BibitemOpen
  \bibfield  {author} {\bibinfo {author} {\bibfnamefont {B.}~\bibnamefont
  {Liang}}, \bibinfo {author} {\bibfnamefont {S.-W.}\ \bibnamefont {Wei}}, \
  and\ \bibinfo {author} {\bibfnamefont {Y.-X.}\ \bibnamefont {Liu}},\ }\href
  {\doibase 10.1142/S0217732319500378} {\bibfield  {journal} {\bibinfo
  {journal} {Mod. Phys. Lett. A}\ }\textbf {\bibinfo {volume} {34}},\ \bibinfo
  {pages} {1950037} (\bibinfo {year} {2019})},\ \Eprint
  {http://arxiv.org/abs/1804.06966} {arXiv:1804.06966 [gr-qc]} \BibitemShut
  {NoStop}%
\bibitem [{\citenamefont {Jiang}(2020)}]{Jiang:2020fgr}%
  \BibitemOpen
  \bibfield  {author} {\bibinfo {author} {\bibfnamefont {S.}~\bibnamefont
  {Jiang}},\ }\href@noop {} {\  (\bibinfo {year} {2020})},\ \Eprint
  {http://arxiv.org/abs/2009.02944} {arXiv:2009.02944 [gr-qc]} \BibitemShut
  {NoStop}%
\bibitem [{\citenamefont {Sorce}\ and\ \citenamefont
  {Wald}(2017)}]{Sorce:2017dst}%
  \BibitemOpen
  \bibfield  {author} {\bibinfo {author} {\bibfnamefont {J.}~\bibnamefont
  {Sorce}}\ and\ \bibinfo {author} {\bibfnamefont {R.~M.}\ \bibnamefont
  {Wald}},\ }\href {\doibase 10.1103/PhysRevD.96.104014} {\bibfield  {journal}
  {\bibinfo  {journal} {Phys. Rev.}\ }\textbf {\bibinfo {volume} {D96}},\
  \bibinfo {pages} {104014} (\bibinfo {year} {2017})},\ \Eprint
  {http://arxiv.org/abs/1707.05862} {arXiv:1707.05862 [gr-qc]} \BibitemShut
  {NoStop}%
%%CITATION = ARXIV:1707.05862;%%
\bibitem [{\citenamefont {Jiang}\ \emph
  {et~al.}(2019{\natexlab{a}})\citenamefont {Jiang}, \citenamefont {Deng},\
  and\ \citenamefont {Chen}}]{Jiang:2019ige}%
  \BibitemOpen
  \bibfield  {author} {\bibinfo {author} {\bibfnamefont {J.}~\bibnamefont
  {Jiang}}, \bibinfo {author} {\bibfnamefont {B.}~\bibnamefont {Deng}}, \ and\
  \bibinfo {author} {\bibfnamefont {Z.}~\bibnamefont {Chen}},\ }\href {\doibase
  10.1103/PhysRevD.100.066024} {\bibfield  {journal} {\bibinfo  {journal}
  {Phys. Rev.}\ }\textbf {\bibinfo {volume} {D100}},\ \bibinfo {pages} {066024}
  (\bibinfo {year} {2019}{\natexlab{a}})},\ \Eprint
  {http://arxiv.org/abs/1909.02219} {arXiv:1909.02219 [hep-th]} \BibitemShut
  {NoStop}%
%%CITATION = ARXIV:1909.02219;%%
\bibitem [{\citenamefont {Jiang}\ \emph
  {et~al.}(2019{\natexlab{b}})\citenamefont {Jiang}, \citenamefont {Liu},\ and\
  \citenamefont {Zhang}}]{Jiang:2019vww}%
  \BibitemOpen
  \bibfield  {author} {\bibinfo {author} {\bibfnamefont {J.}~\bibnamefont
  {Jiang}}, \bibinfo {author} {\bibfnamefont {X.}~\bibnamefont {Liu}}, \ and\
  \bibinfo {author} {\bibfnamefont {M.}~\bibnamefont {Zhang}},\ }\href
  {\doibase 10.1103/PhysRevD.100.084059} {\bibfield  {journal} {\bibinfo
  {journal} {Phys. Rev.}\ }\textbf {\bibinfo {volume} {D100}},\ \bibinfo
  {pages} {084059} (\bibinfo {year} {2019}{\natexlab{b}})},\ \Eprint
  {http://arxiv.org/abs/1910.04060} {arXiv:1910.04060 [hep-th]} \BibitemShut
  {NoStop}%
%%CITATION = ARXIV:1910.04060;%%
\bibitem [{\citenamefont {Jiang}(2019)}]{Jiang:2019soz}%
  \BibitemOpen
  \bibfield  {author} {\bibinfo {author} {\bibfnamefont {J.}~\bibnamefont
  {Jiang}},\ }\href@noop {} {\  (\bibinfo {year} {2019})},\ \Eprint
  {http://arxiv.org/abs/1912.10826} {arXiv:1912.10826 [gr-qc]} \BibitemShut
  {NoStop}%
%%CITATION = ARXIV:1912.10826;%%
\bibitem [{\citenamefont {He}\ and\ \citenamefont {Jiang}(2019)}]{He:2019mqy}%
  \BibitemOpen
  \bibfield  {author} {\bibinfo {author} {\bibfnamefont {Y.-L.}\ \bibnamefont
  {He}}\ and\ \bibinfo {author} {\bibfnamefont {J.}~\bibnamefont {Jiang}},\
  }\href {\doibase 10.1103/PhysRevD.100.124060} {\bibfield  {journal} {\bibinfo
   {journal} {Phys. Rev.}\ }\textbf {\bibinfo {volume} {D100}},\ \bibinfo
  {pages} {124060} (\bibinfo {year} {2019})},\ \Eprint
  {http://arxiv.org/abs/1912.05217} {arXiv:1912.05217 [hep-th]} \BibitemShut
  {NoStop}%
%%CITATION = ARXIV:1912.05217;%%
\bibitem [{\citenamefont {Jiang}\ and\ \citenamefont
  {Zhang}(2020{\natexlab{a}})}]{Jiang:2020alh}%
  \BibitemOpen
  \bibfield  {author} {\bibinfo {author} {\bibfnamefont {J.}~\bibnamefont
  {Jiang}}\ and\ \bibinfo {author} {\bibfnamefont {M.}~\bibnamefont {Zhang}},\
  }\href@noop {} {\  (\bibinfo {year} {2020}{\natexlab{a}})},\ \Eprint
  {http://arxiv.org/abs/2008.04906} {arXiv:2008.04906 [gr-qc]} \BibitemShut
  {NoStop}%
\bibitem [{\citenamefont {Wang}\ and\ \citenamefont
  {Jiang}(2020)}]{Wang:2020vpn}%
  \BibitemOpen
  \bibfield  {author} {\bibinfo {author} {\bibfnamefont {X.-Y.}\ \bibnamefont
  {Wang}}\ and\ \bibinfo {author} {\bibfnamefont {J.}~\bibnamefont {Jiang}},\
  }\href {\doibase 10.1007/JHEP05(2020)161} {\bibfield  {journal} {\bibinfo
  {journal} {JHEP}\ }\textbf {\bibinfo {volume} {05}},\ \bibinfo {pages} {161}
  (\bibinfo {year} {2020})},\ \Eprint {http://arxiv.org/abs/2004.12120}
  {arXiv:2004.12120 [hep-th]} \BibitemShut {NoStop}%
\bibitem [{\citenamefont {Liu}\ and\ \citenamefont {Gao}(2020)}]{Liu:2020cji}%
  \BibitemOpen
  \bibfield  {author} {\bibinfo {author} {\bibfnamefont {C.}~\bibnamefont
  {Liu}}\ and\ \bibinfo {author} {\bibfnamefont {S.}~\bibnamefont {Gao}},\
  }\href {\doibase 10.1103/PhysRevD.101.124067} {\bibfield  {journal} {\bibinfo
   {journal} {Phys. Rev. D}\ }\textbf {\bibinfo {volume} {101}},\ \bibinfo
  {pages} {124067} (\bibinfo {year} {2020})},\ \Eprint
  {http://arxiv.org/abs/2003.12999} {arXiv:2003.12999 [gr-qc]} \BibitemShut
  {NoStop}%
\bibitem [{\citenamefont {Shaymatov}\ \emph {et~al.}(2020)\citenamefont
  {Shaymatov}, \citenamefont {Dadhich}, \citenamefont {Ahmedov},\ and\
  \citenamefont {Jamil}}]{Shaymatov:2019del}%
  \BibitemOpen
  \bibfield  {author} {\bibinfo {author} {\bibfnamefont {S.}~\bibnamefont
  {Shaymatov}}, \bibinfo {author} {\bibfnamefont {N.}~\bibnamefont {Dadhich}},
  \bibinfo {author} {\bibfnamefont {B.}~\bibnamefont {Ahmedov}}, \ and\
  \bibinfo {author} {\bibfnamefont {M.}~\bibnamefont {Jamil}},\ }\href
  {\doibase 10.1140/epjc/s10052-020-8009-4} {\bibfield  {journal} {\bibinfo
  {journal} {Eur. Phys. J. C}\ }\textbf {\bibinfo {volume} {80}},\ \bibinfo
  {pages} {481} (\bibinfo {year} {2020})},\ \Eprint
  {http://arxiv.org/abs/1908.01195} {arXiv:1908.01195 [gr-qc]} \BibitemShut
  {NoStop}%
\bibitem [{\citenamefont {Jiang}\ and\ \citenamefont
  {Zhang}(2020{\natexlab{b}})}]{Jiang:2020btc}%
  \BibitemOpen
  \bibfield  {author} {\bibinfo {author} {\bibfnamefont {J.}~\bibnamefont
  {Jiang}}\ and\ \bibinfo {author} {\bibfnamefont {M.}~\bibnamefont {Zhang}},\
  }\href {\doibase 10.1140/epjc/s10052-020-7751-y} {\bibfield  {journal}
  {\bibinfo  {journal} {Eur. Phys. J. C}\ }\textbf {\bibinfo {volume} {80}},\
  \bibinfo {pages} {196} (\bibinfo {year} {2020}{\natexlab{b}})}\BibitemShut
  {NoStop}%
\bibitem [{\citenamefont {Jiang}\ and\ \citenamefont
  {Gao}(2020)}]{Jiang:2020mws}%
  \BibitemOpen
  \bibfield  {author} {\bibinfo {author} {\bibfnamefont {J.}~\bibnamefont
  {Jiang}}\ and\ \bibinfo {author} {\bibfnamefont {Y.}~\bibnamefont {Gao}},\
  }\href {\doibase 10.1103/PhysRevD.101.084005} {\bibfield  {journal} {\bibinfo
   {journal} {Phys. Rev. D}\ }\textbf {\bibinfo {volume} {101}},\ \bibinfo
  {pages} {084005} (\bibinfo {year} {2020})},\ \Eprint
  {http://arxiv.org/abs/2003.07501} {arXiv:2003.07501 [hep-th]} \BibitemShut
  {NoStop}%
\bibitem [{\citenamefont {Wang}\ and\ \citenamefont
  {Jiang}(2019)}]{Wang:2019bml}%
  \BibitemOpen
  \bibfield  {author} {\bibinfo {author} {\bibfnamefont {X.-Y.}\ \bibnamefont
  {Wang}}\ and\ \bibinfo {author} {\bibfnamefont {J.}~\bibnamefont {Jiang}},\
  }\href@noop {} {\  (\bibinfo {year} {2019})},\ \Eprint
  {http://arxiv.org/abs/1911.03938} {arXiv:1911.03938 [hep-th]} \BibitemShut
  {NoStop}%
%%CITATION = ARXIV:1911.03938;%%
\bibitem [{\citenamefont {Zeng}\ \emph
  {et~al.}(2019{\natexlab{b}})\citenamefont {Zeng}, \citenamefont {Hu},\ and\
  \citenamefont {He}}]{Zeng:2019hux}%
  \BibitemOpen
  \bibfield  {author} {\bibinfo {author} {\bibfnamefont {X.-X.}\ \bibnamefont
  {Zeng}}, \bibinfo {author} {\bibfnamefont {X.-Y.}\ \bibnamefont {Hu}}, \ and\
  \bibinfo {author} {\bibfnamefont {K.-J.}\ \bibnamefont {He}},\ }\href@noop {}
  {\bibfield  {journal} {\bibinfo  {journal} {Nucl. Phys.}\ }\textbf {\bibinfo
  {volume} {B949}},\ \bibinfo {pages} {114823} (\bibinfo {year}
  {2019}{\natexlab{b}})},\ \Eprint {http://arxiv.org/abs/1905.07750}
  {arXiv:1905.07750 [hep-th]} \BibitemShut {NoStop}%
%%CITATION = ARXIV:1905.07750;%%
\bibitem [{\citenamefont {Ge}\ \emph {et~al.}(2018)\citenamefont {Ge},
  \citenamefont {Mo}, \citenamefont {Zhao},\ and\ \citenamefont
  {Zheng}}]{Ge:2017vun}%
  \BibitemOpen
  \bibfield  {author} {\bibinfo {author} {\bibfnamefont {B.}~\bibnamefont
  {Ge}}, \bibinfo {author} {\bibfnamefont {Y.}~\bibnamefont {Mo}}, \bibinfo
  {author} {\bibfnamefont {S.}~\bibnamefont {Zhao}}, \ and\ \bibinfo {author}
  {\bibfnamefont {J.}~\bibnamefont {Zheng}},\ }\href {\doibase
  10.1016/j.physletb.2018.07.015} {\bibfield  {journal} {\bibinfo  {journal}
  {Phys. Lett.}\ }\textbf {\bibinfo {volume} {B783}},\ \bibinfo {pages} {440}
  (\bibinfo {year} {2018})},\ \Eprint {http://arxiv.org/abs/1712.07342}
  {arXiv:1712.07342 [hep-th]} \BibitemShut {NoStop}%
%%CITATION = ARXIV:1712.07342;%%
\bibitem [{\citenamefont {Iyer}\ and\ \citenamefont
  {Wald}(1995)}]{Iyer:1995kg}%
  \BibitemOpen
  \bibfield  {author} {\bibinfo {author} {\bibfnamefont {V.}~\bibnamefont
  {Iyer}}\ and\ \bibinfo {author} {\bibfnamefont {R.~M.}\ \bibnamefont
  {Wald}},\ }\href {\doibase 10.1103/PhysRevD.52.4430} {\bibfield  {journal}
  {\bibinfo  {journal} {Phys. Rev.}\ }\textbf {\bibinfo {volume} {D52}},\
  \bibinfo {pages} {4430} (\bibinfo {year} {1995})},\ \Eprint
  {http://arxiv.org/abs/gr-qc/9503052} {arXiv:gr-qc/9503052 [gr-qc]}
  \BibitemShut {NoStop}%
%%CITATION = GR-QC/9503052;%%
\bibitem [{\citenamefont {Figueras}\ \emph {et~al.}(2017)\citenamefont
  {Figueras}, \citenamefont {Kunesch}, \citenamefont {Lehner},\ and\
  \citenamefont {Tunyasuvunakool}}]{Figueras:2017zwa}%
  \BibitemOpen
  \bibfield  {author} {\bibinfo {author} {\bibfnamefont {P.}~\bibnamefont
  {Figueras}}, \bibinfo {author} {\bibfnamefont {M.}~\bibnamefont {Kunesch}},
  \bibinfo {author} {\bibfnamefont {L.}~\bibnamefont {Lehner}}, \ and\ \bibinfo
  {author} {\bibfnamefont {S.}~\bibnamefont {Tunyasuvunakool}},\ }\href
  {\doibase 10.1103/PhysRevLett.118.151103} {\bibfield  {journal} {\bibinfo
  {journal} {Phys. Rev. Lett.}\ }\textbf {\bibinfo {volume} {118}},\ \bibinfo
  {pages} {151103} (\bibinfo {year} {2017})},\ \Eprint
  {http://arxiv.org/abs/1702.01755} {arXiv:1702.01755 [hep-th]} \BibitemShut
  {NoStop}%
%%CITATION = ARXIV:1702.01755;%%
\bibitem [{\citenamefont {Kubiznak}\ and\ \citenamefont
  {Mann}(2012)}]{Kubiznak:2012wp}%
  \BibitemOpen
  \bibfield  {author} {\bibinfo {author} {\bibfnamefont {D.}~\bibnamefont
  {Kubiznak}}\ and\ \bibinfo {author} {\bibfnamefont {R.~B.}\ \bibnamefont
  {Mann}},\ }\href {\doibase 10.1007/JHEP07(2012)033} {\bibfield  {journal}
  {\bibinfo  {journal} {JHEP}\ }\textbf {\bibinfo {volume} {07}},\ \bibinfo
  {pages} {033} (\bibinfo {year} {2012})},\ \Eprint
  {http://arxiv.org/abs/1205.0559} {arXiv:1205.0559 [hep-th]} \BibitemShut
  {NoStop}%
\end{thebibliography}
\end{document}